\documentclass[twocolumn]{aastex631}

\accepted{August 27, 2024}

\submitjournal{ApJ Letters}

\shorttitle{Radius Distribution Depends on Gap Complexity}
\shortauthors{Rice et al.}
\usepackage{ulem,bm}

\begin{document}

\title{The Distribution of Planet Radius in Kepler Multiplanet Systems Depends on Gap Complexity}

\correspondingauthor{David R. Rice}
\email{davidr@post.openu.ac.il}

\author[0000-0001-6009-8685]{David R. Rice}
\affiliation{Astrophysics Research Center (ARCO), Department of Natural Sciences, The Open University of Israel, Raanana 4353701, Israel}

\author[0000-0003-2202-3847]{Jason H. Steffen}
\affiliation{Department of Physics and Astronomy, University of Nevada, Las Vegas, 4505 South Maryland Parkway, Las Vegas, NV 89154, USA}
\affiliation{Nevada Center for Astrophysics, University of Nevada, Las Vegas, 4505 South Maryland Parkway, Las Vegas, NV 89154, USA}

\author[0000-0001-9504-3174]{Allona Vazan}
\affiliation{Astrophysics Research Center (ARCO), Department of Natural Sciences, The Open University of Israel, Raanana 4353701, Israel}



\begin{abstract}

The distribution of small planet radius ($<$4 R$_\oplus$) is an indicator of the underlying processes governing planet formation and evolution. We investigate the correlation between the radius distribution of exoplanets in \textit{Kepler} multiplanet systems and the system-level complexity in orbital period spacing. Utilizing a sample of 234 planetary systems with three or more candidate planets orbiting FGK main-sequence stars, we measure the gap complexity ($C$) to characterize the regularity of planetary spacing and compare it with other measures of period spacing and spacing uniformity. We find that systems with higher gap complexity exhibit a distinct radius distribution compared to systems with lower gap complexity. Specifically, we find that the radius valley, which separates super-Earths and sub-Neptunes, is more pronounced in systems with lower gap complexity ($C$$<$0.165). Planets in high complexity systems ($C$$>$0.35) exhibit a lower frequency of sub-Earths (2.5 times less) and sub-Neptunes (1.3 times less) and a higher frequency of super-Earths (1.4 times more) than planets in low complexity systems. This may suggest that planetary systems with more irregular spacings are more likely to undergo dynamic interactions that influence planet scattering, composition, and atmospheric retention. The gap complexity metric proves to be a valuable tool in linking the orbital configurations of planets to their physical characteristics.

\end{abstract}



\section{Introduction} \label{sec:intro}

Around stars like the Sun on orbits shorter than than one year, the most common size of planet is between the radius of Earth and the radius of Uranus (1-4 R$_\oplus$). \citet{Borucki2011} and \citet{Batalha2013} first identified this trend in planets observed by the \textit{Kepler} space telescope. Occurrence rate studies (e.g. \citealt{Howard2012,Petigura2013, Burke2015}) which correct for the biases of observing transiting planets further confirm the overabundance of planets larger than the Earth but smaller than a gas-rich planet (5-10\% gas by mass).

However, the distribution of planet sizes is not uniform over the interval of 1-4 R$_\oplus$. \citet{Fulton2017} finds a deficit of planets with radii between 1.5 and 2.0 R$_\oplus$. The bi-modality in small planet radii separated by this radius ``valley'' may indicate a bi-modality in the volatile mass of planets. Planets below the radius valley will be referred to as ``super-Earths'' in this work and cannot be composed of more than 50\% water \citep{Zeng2019, Unterborn2023}. While planets above the valley are referred to as ``sub-Neptunes'' and require large water fractions or a gas atmosphere.

From just observing planet radii, constraints can be placed on the formation and evolution of planets between 1-4 R$_\oplus$. Small planets may either form from a bimodal amount of water \citep{Luque2022,Burn2024} or planets in the radius valley must undergo the loss of their volatiles \citep{Owen2013, Chen2016}. In the later case, planets above the valley are massive enough to hold onto light gasses, while planets in the valley lose gasses when stripped by high energy photons \citep{Owen2017, VanEylen2018, Rogers2021} and when internally heated by both core formation \citep{Ginzburg2018, Gupta2019, Gupta2020} and giant impacts \citep{Biersteker2019}.

The radius valley is observed for planets on periods shorter than 100 days around FGK stars and is sensitive to the formation environment of planets. The valley can shift based on stellar temperature, the orbital period of planets, the composition of the planets and their building blocks, and the distribution of planet mass \citep{Gupta2019}. The radius valley is shown to be shallower around low mass stars in \citet{Ho2024}. From \citet{Sullivan2023}, the radius valley may be non-existent for planets in binary star systems. The radius distribution along with the mass distribution are compared for single planets verses multiple planet systems in works such as \citet{Latham2011}, \citet{RodriguezMartinez2023multi}, and \citet{Liberles2023}. Understanding the radius distribution's relationship to other system parameters may disentangle the formation mechanisms responsible for the radius valley.

\textit{Kepler} also gave us a wealth of information on systems with multiple transiting planets. We will shorten multiple planet systems to ``multis'' in this work. Multis tend to host planets closely packed together with the ratio of adjacent planet's periods smaller than that of the Solar System's \citep{fang}. Long chains of orbital resonances stabilize some multis (e.g. \citealt{Mills2016,Luger2017,MacDonald2022}). \citet{Weiss2018} showed that multis tend to have planets of similar sizes and regular spacings, analogous to the sizes and spacings of peas in a pea pod. This trend of uniform properties within a system also applies to planet mass \citep{Millholland2017}. 

The ``peas in a pod'' observations constrain potential models for the formation of planetary systems.  Plausible models to produce the intra-system uniformity in period spacing include the redistribution of planets via dynamical instability \citep{Goldberg2022, Lammers2023, Ghosh2024}, migration within the protoplanetary disk \citep{Batygin2023b,Goldberg2023,Choksi2023,Shariat2024}, and the minimization of energy within a low-mass system of planets \citep{Adams2020}.

The mechanism that lead to the regular spacing of planets in multis may also affect the composition of the planets formed. In this work, we examine the frequency of planets near the radius valley in multis with uniform and irregular spacings. To summarize the spacing of planets in multis, we calculate a system's gap complexity which only depends on planet period. If both the complexity and planet radius frequency change in tandem with one another we may be able to tie together formation models.

A large motivator for this work is \citet{He2023} who found a correlation between gap complexity in the inner systems and the existence of an outer, cold giant planet (50 M$_\oplus$$\leq$M$_p \sin i$$\le$13 M$_{Jup}$). Using the \textit{Kepler} Giant Planet Search \citep{Weiss2024}, they found that all four systems with 3+ transiting inner planets and an outer giant planet are irregularly spaced with gap complexities above 0.32. \citet{Sullivan2023} and \citet{Sullivan2024} show that stellar companions affect the radius distribution, hence outer giant planets may similarly affect the formation environments of inner systems and the radius distribution of small planets.

Our paper is laid out as follows. In Sect.~\ref{sec:systems}, we define our sample of planet candidates in multis around FGK main-sequence stars observed by \textit{Kepler}. We find the gap complexity of these systems and report any correlations with underlying variables in Sec.~\ref{sec:gap}. In Sec.~\ref{sec:radius}, we show how the planetary radius distribution changes in systems of increasing gap complexity and changes when splitting systems into low and high gap complexity (Sec.~\ref{sec:radsplit}). We discuss changes in the radius distribution in Sec.~\ref{sec:discussion} and conclude in Sec.~\ref{sec:conclusion}. 

\section{System Selection} \label{sec:systems}

The recent catalog of planet candidates observed by the \textit{Kepler} space telescope, \citet{Lissauer2023}, found 253 planetary systems of three or more candidate planets. One focus of this catalogue which contains 4376 candidates and 709 multis is providing accurate planet parameters in multis. The radius distribution is likely different around low-mass stars \citep{Morton2014, VanEylen2021,Ho2024} and the catalogue has few evolved stars. Thus, we limit systems to FGK main-sequence stars between 4000 and 7000 K which removes 12 systems and with surface gravity over 10$^4$ which removes an additional three systems. These cuts are similar to those in \citet{Hsu2018}.

In this work we aim to study the gap complexity in the inner planets of a system. Thus, we remove planets with orbital periods over 1000 days. This removes three of the longest period planets while keeping the systems as complete as possible for the calculation of gap complexity.

\begin{figure}[ht!]
\plotone{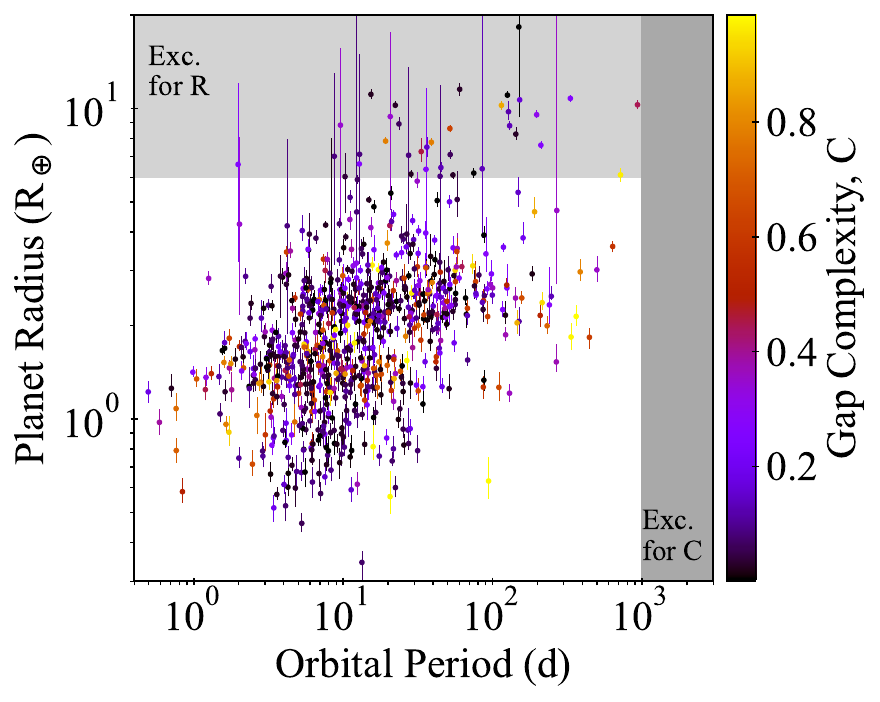}
\caption{Planet radius with radius uncertainties verses orbital period  for planets in systems of more than 3-planets. We use the data from \citet{Lissauer2023} and limit the study to FGK stars. The radius gap is visible near 2 R$_\oplus$ with a slope which depends on orbital period. Planets with periods over 1000 days are excluded from the gap complexity ($C$) analysis. Additionally, planets with a radius over 6 $R_\oplus$ or radius uncertainties over 50\% are used in the $C$ calculation but excluded from the radius distribution analysis.
\label{fig:allpl}}
\end{figure}

Our sample of multis includes 825 candidate planets in 238 planetary systems which have three or more planets. The periods and radii of the planets are shown in Fig.~\ref{fig:allpl}. The maximum error in the planet period is 0.04\% with a median error below 0.001\%. A majority of our 238 systems are 3-planet systems making up 68.5\% of the systems. 20.2\% are in 4-planet systems. 11.3\% are in higher multiplicity systems (21 in 5-planet, 4 in 6-planet, 1 in 7-planet, and 1 in 8-planet systems).

Other works have more stringent limits on the maximum orbital period. \citet{He2023} limits systems to one year periods. This limit would remove 15 planets and would result in three systems no longer having three planets. The completeness of the \textit{Kepler} survey begins to drop at periods of 100 days. A 100 day cut removes 12\% of planets and results in 16\% of systems (38) no longer having three planets compared to our 1000 day cut. While not presented in this work, our findings on the changes in the radius distribution with gap complexity (Sec.~\ref{sec:radius}) hold for a 100 day period cutoff with only slightly less significance primarily from the number of systems decreasing by 16\%.

\section{Gap Complexity of \textit{Kepler} Multis} \label{sec:gap}

The stability time of a planetary system strongly depends on how far planets are spaced from one another within the system \citep{fang,chambers,obertas,Rice2018}. However, for dynamical considerations spacing is generally measured in multiples of mutual Hill Radii which depends on planet mass. In this work with \textit{Kepler} planets, we have well determined periods but limited mass measurements. Additionally, while many \textit{Kepler} systems are near commensurability with mean motion resonance, the majority are not in resonance \citep{fabrycky,steffen}, so we do not measure planet spacing based upon distance to a mean motion resonance. Thus, we adopt the gap complexity as a measure to describe the architecture of multis. 

Gap complexity, $C$, measures how evenly spaced planets are in log-period space for systems of three or more planets. Gap complexity was introduced in \citet{Gilbert2020} following the work of \citet{LOPEZRUIZ1995} which defined complexity as the product of information and disequilibrium in a system. For planet periods, \citet{Gilbert2020} define gap complexity as
\begin{equation}\label{eq:C}
C=-K\left(\sum^{n-1}_{i=1}p_i^*\ln p_i^*\right)\left(\sum^{n-1}_{i=1}\left(p_i^*-\frac{1}{n-1}\right)^2\right),
\end{equation}

\begin{equation}
p_i^*=\frac{\ln(P_{i+1}/P_i)}{\ln(P_{n}/P_{1})},
\end{equation}
where n is the number of planets in the system, the ratio of periods for each adjacent planet pair is $P_{i+1}/P_{i}$, and the outermost and innermost planet periods are $P_{n}$ and $P_{1}$. 

In Eq.~\ref{eq:C}, $K$ is a normalization constant which varies based on planet multiplicity and given in \citet{Gilbert2020} as $K=1/C_{max}$ where $C_{max}$=[0.105, 0.212, 0.291, 0.350, 0.398, 0.437] respectively for each multiplicity from 3- to 8-planet. 

This normalization makes $C$ vary for all planet multiplicity from 0, being evenly spaced in log-period, and 1, being the max complexity. For the Solar System, the inner four planets have a gap complexity of 0.126 and removing Mercury results in a low gap complexity of 0.055. 

\begin{figure}[ht!]
\plotone{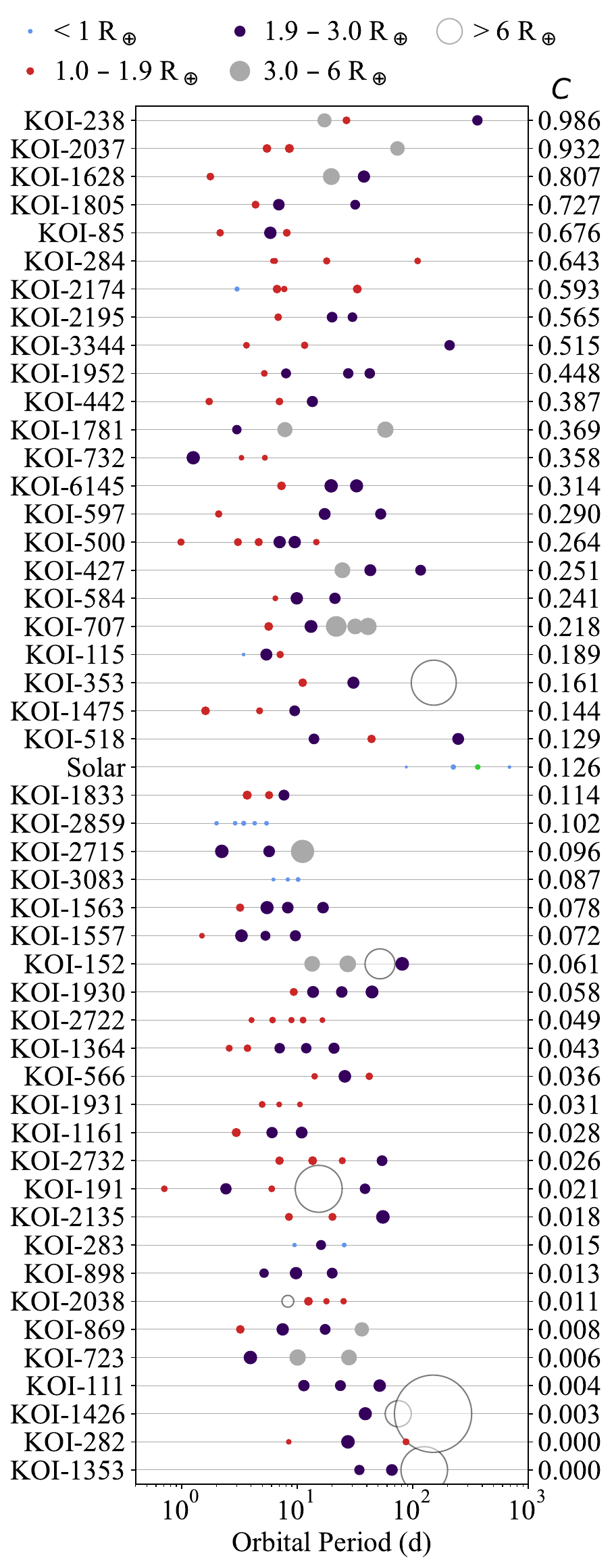}
\caption{Architectures of planetary systems where each horizontal line is a system with the star's KOI number and the system's $C$ noted. Every fifth planetary system when the 238 systems are ordered by $C$ (48 systems) are shown.  Points are placed for each planet at its orbital period with a size proportional to the planet radius. Planets are colored by size category. Unfilled planets are over 6 R$_\oplus$ or have radius uncertainties over 50\%. Unfilled planets are used to calculate the gap complexity, but the radius is not used in later analyses. The inner Solar System is also shown.
\label{fig:arch}}
\end{figure}

In Fig.~\ref{fig:arch}, we plot the orbital architectures of 20\% of the systems in our sample ordered by their gap complexity. The systems are chosen so that the proportions of systems below/above a given $C$ is representative of the entire sample. As $C$ increases the spacing between planets in log-period space becomes visually more diverse. 

Additionally in Fig.~\ref{fig:periods}, we show an empirical cumulative distribution function of gap complexity for our systems. The distribution is in agreement with the DR25 catalogue of \textit{Kepler} candidate planets (\citealt{koidr25}, also analyzed in \citet{He2023}). With our relaxed orbital period cutoff of 1000 days a few systems with low complexity in shorter period planets move to the highest complexities when the exterior planet is included. This is visible as a slight uptick in systems with $C$$>$0.9.

A majority of \textit{Kepler} multis have gap complexities under that of the inner Solar System (0.126). Half of systems are below $C$=0.1057 and 13\% of systems are below $C$=0.01. While the 10\% of systems with highest $C$ spread between $C$=0.66 and the maximum of 0.986 in the Kepler-123 (KOI-238) 3-planet system.

\subsection{Correlations with Other System Properties}

We find that the gap complexity of a system does not correlate with the host star's properties. The sample Pearson correlation coefficients, $r$, between $C$ and stellar temperature, mass, radius, log(g), and metallicity are all below 0.1 with slopes consistent with zero. The average radius uncertainty and the average log-flux of planets in a system has no correlation with gap complexity. Additionally, we find sensitivity-related parameters such as the Multiple Event Statistic (MES) have no correlation with gap complexity both when all values for the planets are used and when values are averaged in a system. We also find the distributions of MES in low and high complexity systems are consistent with each other.

However, the gap complexity is sensitive to other period-related measures. While not correlated with inner period, the outer period has a correlation coefficient of 0.37 with $C$. From a slope, $b$, of a linear regression ($y=a+bx$), the outer planet period around a $C$=1.0 system is expected to be 155 days longer than a $C$=0 system. 

%This correlation still exists if the maximum period allowed for a system is set to 100 days (the slope of outer period with $C$ is then 20 days).

\begin{figure}[ht!]
\plotone{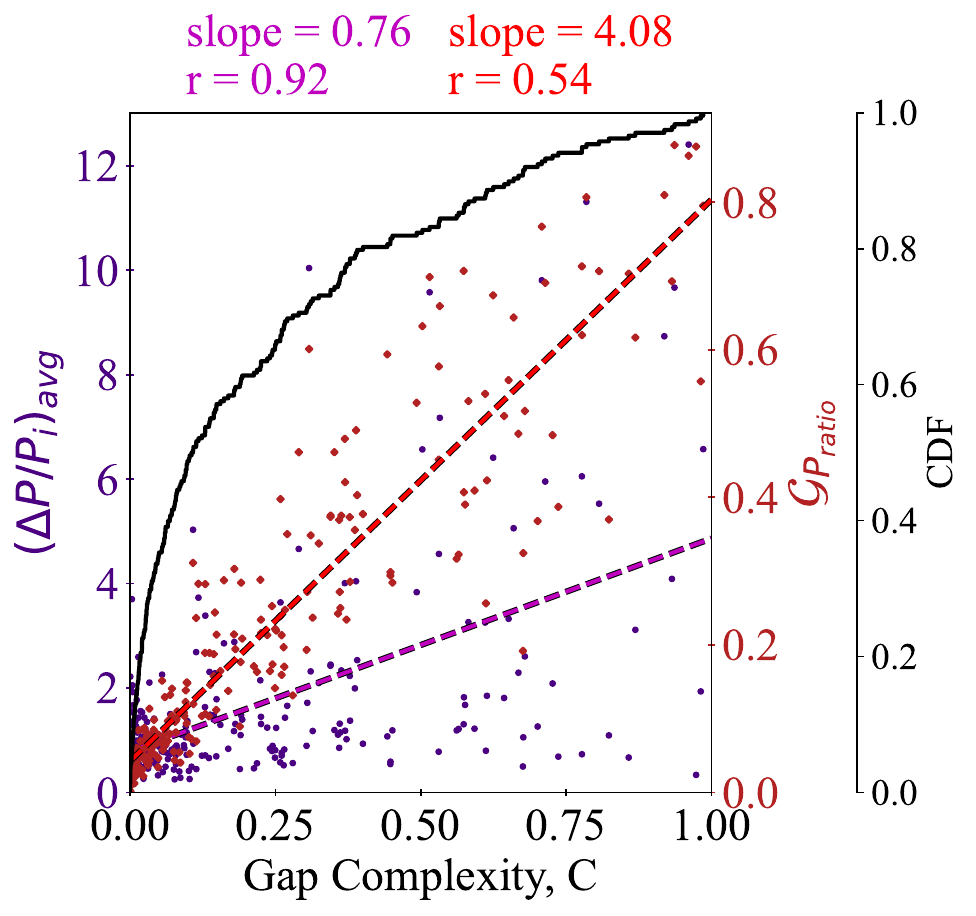}
\caption{\textit{Black}, empirical cumulative distribution function of gap complexity for our 238 systems. \textit{Purple circles}, the average $\Delta P/P_i$ in the systems (left axis) against their gap complexities. Low-$C$ systems have planets close together in period while the dispersion in period spacing is much larger in high-$C$ systems. \textit{Red diamonds}, the Gini index of planet period ratio (right axis) verses gap complexity for our systems. These two measures of period spacing uniformity are highly correlated. For both the \textit{purple} and \textit{red} points a linear regression fit is shown with the slope and correlation coefficient ($r$) reported.
\label{fig:periods}}
\end{figure}

Correlations also exist between gap complexity and the range of periods in log-space ($b=0.72$ days, $r=0.47$) and the normalized standard deviation of period ($\sigma_P/\mu_P$, $b=0.53$, $r=0.49$). The strongest correlation, $r=0.64$, is shown in Fig.~\ref{fig:periods} \textit{left axis} between gap complexity and the average $\Delta P/P_i$ in a system. Where the average $\Delta P/P_i$ is equal to 
\begin{equation}\label{dPP}
\overline{\frac{\Delta P}{P_i}}= \frac{1}{n}\sum^{n-1}_{i=1}\frac{P_{i+1}-P_i}{P_i}.
\end{equation}
Since the ratio of the outer planet to inner planet period in each nearest pair is equal to $1+\Delta P/P_i$, the figure shows a majority of systems with C$<$0.1 have average period ratios less than 3:1. While systems at large gap complexities may have low period ratios, they have a wider spread in period ratios with some over 10:1. 
%These correlations remain significant if we limit periods to 100 days with slightly lower slopes and correlation coefficients.

We find that systems with larger gap complexities tend to be more widely spread out in planet periods. This has not been considered in previous works and for our purposes is considered as an underlying variable in the subsequent sections.

Additionally, other works use different measures to describe how planets are spaced in a system. Following upon the work of \citet{Goyal2022} and \citet{Goyal2023}, \citet{Goyal2024} describes the spacing uniformity in systems with the adjusted Gini index \citep{Deltas2003, Gini1912} of a system's period ratios ($\mathcal{G}_{P_{ratio}}$). The Gini index is an economic measure of inequality and is used here to describe the uniformity of period ratios in a system. In Fig.~\ref{fig:periods} \textit{right axis}, we show the correlation between gap complexity and the Gini index of period ratios. The Gini index is also normalized to be between zero and one, but the maximum in our systems is 0.88. While the correlation is not 1:1 the two parameters to measure the uniformity of spacing in systems are highly correlated.

\section{Changes in the Radius Distribution with Gap Complexity}\label{sec:radius}

We calculate gap complexity using all the planets in our system sample with orbital periods less than 1000 days, but now we limit our radius analysis to planets under 6 R$_\oplus$ and with radius uncertainties below 50\%. This limits the analysis to 783 planets in 234 systems. Four systems go from having three planets to no planets which pass this criteria. Fourteen systems have less than three planets, 151 systems have three planets, and 69 systems have more than three planets. The gap complexity of the four systems with no planets which pass our criteria are $C$=0.08, 0.13, 0.26, 0.34.

\begin{figure}[ht!]
\plotone{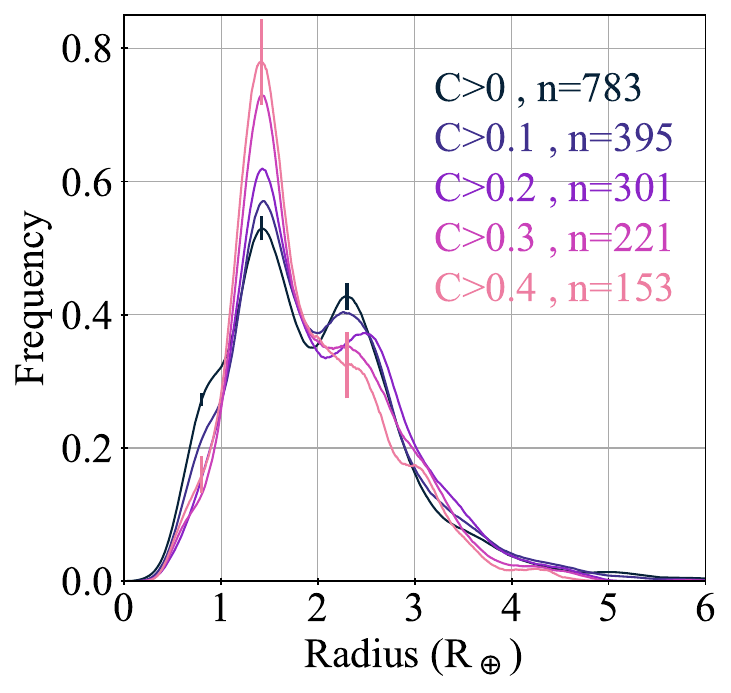}
\caption{The median frequency of planet radius from 500 Gaussian kernel density estimations where each planet's radius is drawn from a normal distribution centered at the planet's reported radius and standard deviation set the average of the plus and minus uncertainties. This is then repeated for each set of planets limited by minimum gap complexity given by the color. The 1$\sigma$ uncertainty is given at three points for the entire set of planets and planets with $C$$>$0.4. While the uncertainty increases from the smaller sample size, there is a systematic increase in super-Earth and decrease in sub-Neptune frequency.  
\label{fig:changing}}
\end{figure}

In Fig.~\ref{fig:changing}, we show how the radius distribution changes for systems with a gap complexity above a given value. All 783 planets are shown by the dark distribution which features a prominent radius valley. The distribution is smoothed over with a Gaussian kernel density estimation (KDE) with a bandwidth equal to the average radius uncertainty. 500 realizations of this KDE is calculated from draws of the 783 planet radii considering their individual uncertainties, and the median of these realizations is shown. The minimum gap complexity is then varied and the distribution recalculated.

As gap complexity increases, we find a decrease in sub-Earth frequency, an increase in super-Earth frequency, and a decrease in sub-Neptune frequency. While the uncertainty grows from the smaller sample size at high-$C$, these features monotonically increase/decrease with minimum gap complexity. The frequency of sub-Neptunes decreases to an equal or lower frequency than planets in the radius valley at large gap complexity.

\begin{figure}[ht!]
\plotone{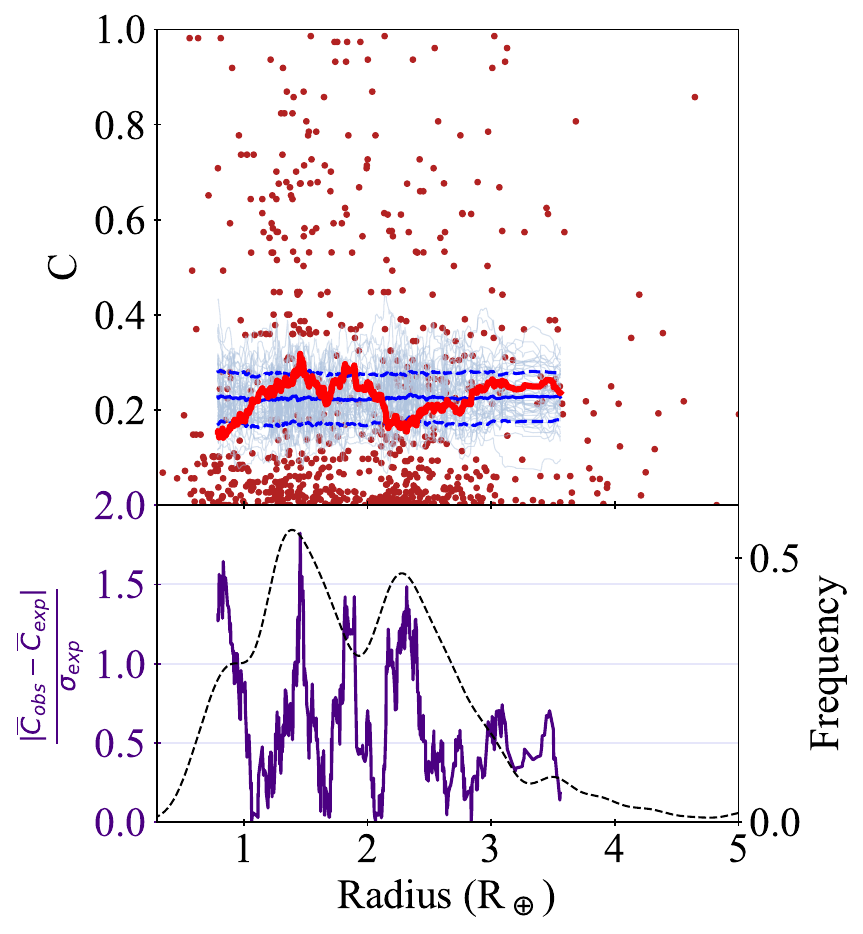}
\caption{\textit{Top}, gap complexity verses planet radius for our 783 planets with a running average of gap complexity in 80 planets (\textit{red line}). \textit{Light blue lines} are 50 realizations of the running average gap complexity if the gap complexities are randomly assigned to systems. \textit{Blue} shows the average and $\pm 1\sigma$ of 500 realizations. \textit{Bottom}, the absolute difference between the average gap complexity in the observations (\textit{red line}) and the average gap complexity in the random shuffled experiment (\textit{blue line}) as a multiple of the standard deviation of the randomized planets. While only between 1-2$\sigma$, the peaks align with peaks and dips in the radius distribution (\textit{black dashed}), planets in these peaks/dips tend to be anomalous in average gap complexity.
\label{fig:sigplot}}
\end{figure}

As a first test of the significance of these features, we test if a given radius prefers systems of higher or lower $C$ than the median $C$. In Fig.~\ref{fig:sigplot} the thick red line shows the running median $C$ of 80 planets when the planets are ordered by their observed median radius. We then create test samples where we randomly assign the gap complexities in our sample to each system. Each thin blue line in Fig.~\ref{fig:sigplot}, is the running median $C$ in a test sample where the gap complexity has been randomly assigned. The median $C$ across all radii for 500 of these test samples is 0.22$\pm$0.05. 

By comparing the median and standard deviation of $C$ in our test samples to the observed $C$, we can understand at what point the radius distribution prefers higher or lower gap complexity. The deviations are only up to 1.75$\sigma$, but they occur at interesting locations in the radius distribution (Fig.~\ref{fig:sigplot}, \textit{bottom}). We see the gap complexity is lower than the expected value in both the sub-Earth regime and sub-Neptune regime. The gap complexity is higher than expected in the super-Earth and radius valley regime. While our test samples often have deviations as large as the observations, they have equal likelihood of appearing anywhere as evidenced by the median and standard deviation being constant across radius.\\

\subsection{Radii in Systems of low- and high-C} \label{sec:radsplit}

In this section, we aim to examine the radius distribution in systems split into samples of low- and high-$C$. First, we determine at what gap complexity to divide the systems. We perform a two-sample Kolmogorov–Smirnov (K-S) test between samples of planet radius divided into low-/high-$C$ with various maximum/minimum values of $C$ determining the samples. A two-sample K-S test gives the probability that both samples were drawn from the same underlying population without assumption of that underlying distribution. We emphasize that the K-S test is sensitive to sample size and cannot be used to tell how different or in what ways two distributions are different.  For example, a p-value less than 0.05 can be achieved by drawing 30 values from $\mathcal{N}(0,1)$ and from $\mathcal{N}(1,1)$ or similarly by drawing 3000 values from $\mathcal{N}(0,1)$ and from $\mathcal{N}(0,0.9)$.

\begin{figure}[ht!]
\plotone{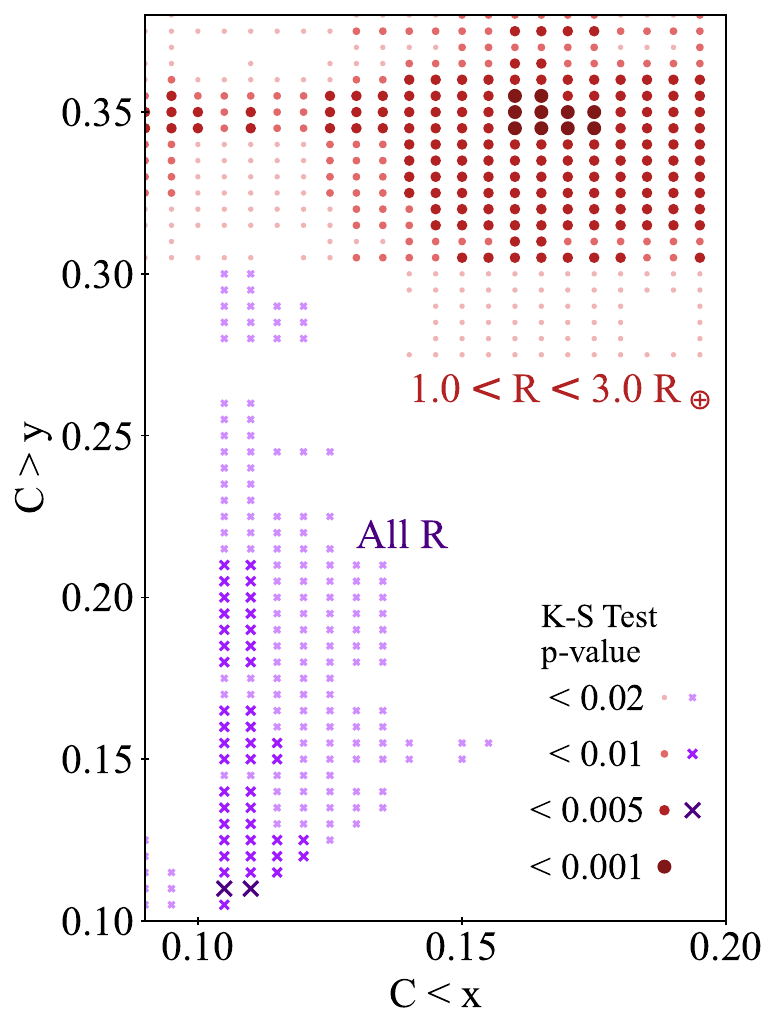}
\caption{We perform K-S tests between the radius of planets separated by the system's gap complexities ($C$). One radius sample has a maximum $C$ given by the x-axis and the other has a minimum set by the y-axis. The significance when using all planets from 0-6 R$_\oplus$ is shown with \textit{purple x's}. While the significance for planets between 1-3 R$_\oplus$ is shown with \textit{red circles}. Size and darkness of the marker represents the significance value without considering radius uncertainties. 
\label{fig:kstests}}
\end{figure}

In Fig.~\ref{fig:kstests}, we show the p-value of the K-S test with different divisions. When using all planet radii ($<$6 R$_\oplus$), the most significance occurs when the sample is divided into systems with $C$$<$0.105 and systems of $C$$>$0.11. However, our primarily interest is in in the region between one and three Earth-radii which features the radius valley. Thus, we also perform K-S tests only considering planets in this region. The significance comes when low-$C$ is defined as $C$$<$0.165 and high-$C$ is defined as $C$$>$0.35.

Moving forward we will divide out samples into low- and high-$C$ based upon the minimum p-value of the K-S tests since no division exists a priori. However, Fig.~\ref{fig:kstests} shows these points are not exceptional with significance varying slowly and smoothly in most directions from the chosen division. When only considering the planets with 1$<$R$<$3 R$_\oplus$, there is significance across all cuts of low-$C$ for any high-$C$ cut above 0.3.

\begin{figure*}[ht!]
\plottwo{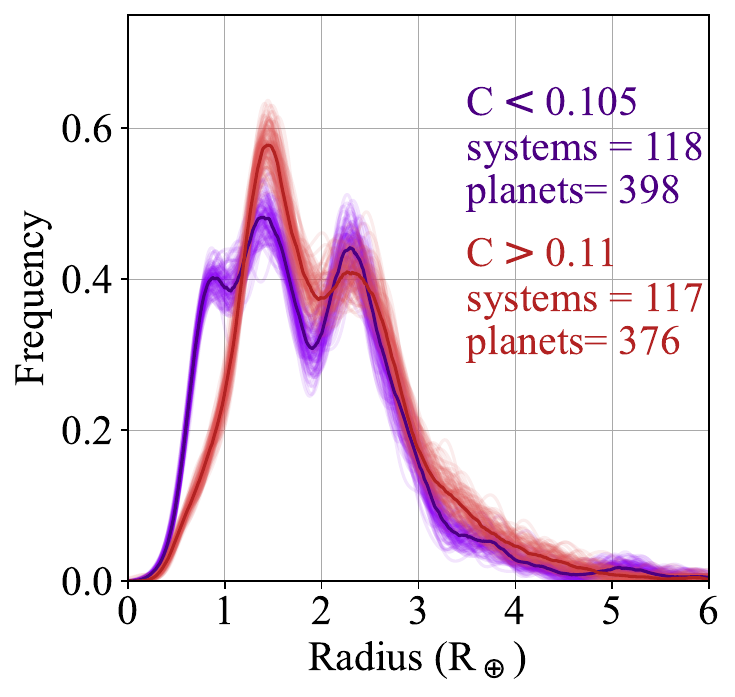}{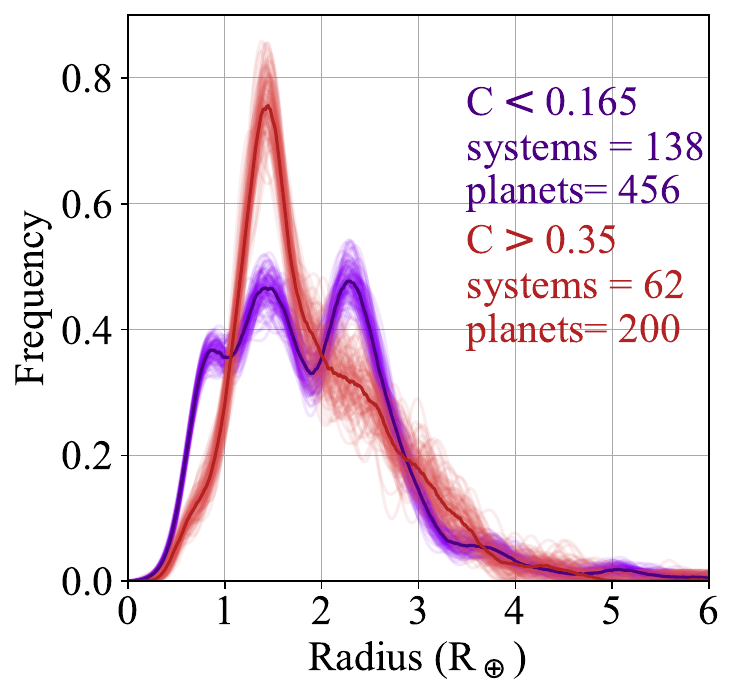}
\caption{The radius distribution at low and high gap complexity with divisions based on the K-S tests from Fig.~\ref{fig:kstests}. For each sample, 100 Gaussian kdes of radius draws from reported observational uncertainties with bandwidth equal to the average radius error. Thick line shows the average of the 100 realizations. \textit{Left}, radius of planets split by their system's $C$ for the minimum p-value of a K-S test across all radii. \textit{Right}, radius of planets split by their system's $C$ for the minimum p-value of a K-S test in the 1-3 R$_\oplus$ region.  
\label{fig:radiisplit}}
\end{figure*}

In Fig.~\ref{fig:radiisplit}, we show the radius distribution for systems below and above the determined gap complexities.  For each of the low- and high-$C$ systems, we show 100 distributions where each planet's radius is randomly chosen from a Gaussian with observed median value and a standard deviation equal to the average of the plus/minus observed uncertainties. We calculate the p-value of a 2-sample K-S test between each sampled radius distribution for the 100 distributions. 

When considering all planets, there are 118 systems with $C$$<$0.105 and 117 systems with $C$$>$0.11 and 3 systems between. The median p-value of a 2-sample K-S test between these two distributions is 0.0044. We see when considering all planets, Fig.~\ref{fig:radiisplit} \textit{left}, the abundance of sub-Earths at low-$C$ dominate the significance.

In Fig.~\ref{fig:radiisplit} \textit{right}, we show the entire radius distribution for the highest significance in the 1-3 R$_\oplus$ region. There are 138 systems with $C$$<$0.165 and 62 systems with $C$$>$0.35. The median p-value of a 2-sample K-S test between the planets in the analysis region of 1-3 R$_\oplus$ is 0.0008. We see the frequency of sub-Neptunes in the high-$C$ systems has been suppressed to below the frequency of planets in the radius valley. The frequency of super-Earths increases from the low-$C$ to the high-$C$.

For the region of 1-3 R$_\oplus$, we also perform a test of modality on the low- and high-$C$ planets. Using a Hartigan dip-test of unimodality at a p-value of 0.01 and 1000 trials we find the low-$C$ planets to have two peaks while the high-$C$ only has one peak. The test find the two peaks to be from 1.08 to 1.9 R$_\oplus$ and from 1.90 to 2.40 R$_\oplus$. While for the high-gap complexity planets, the test returns one broad peak across the analysis range from 1.01 to 2.98 R$_\oplus$.

\subsection{Sub-Earth Abundance} \label{sec:subearth}

In our entire sample, 47 systems have 95 planets with median planet radius less than the Earth. Systems with sub-Earths span gap complexities from 0-1. However, we see in Fig.~\ref{fig:sigplot} sub-Earths have a 1.5$\sigma$ lower gap complexity than expected. In Fig.~\ref{fig:changing}, we see a bump in the sub-Earth frequency at low-$C$ which decreases at gap complexities over 0.2 to the frequency consistent with a tail of a Gaussian which is centered in the super-Earth region.

When dividing based on the K-S test in Fig.~\ref{fig:kstests}, 27 of the 47 sub-Earth hosting systems have $C$$<$0.105 and host 67 sub-Earths. While, 19 sub-Earth systems have $C$$>$0.11 and host 27 sub-Earths. At this division there are approximately the same amount of total systems and planets in the low- and high-$C$ samples (Fig.~\ref{fig:radiisplit}). However, low-$C$ systems host approximately 2.5 times more sub-Earths than high-$C$ systems. 

\subsection{Super-Earth and Sub-Neptune Abundance} \label{sec:radgap}

The minimum frequency of planet radius in the region of 1-3 R$_\oplus$, the bottom of the radius valley, occurs at approximately 1.9 R$_\oplus$ in our total planet sample. We will define super-Earths as planets larger than Earth and smaller than 1.9 R$_\oplus$. Sub-Neptunes will be shorthand for those above 1.9 R$_\oplus$ and below 3 R$_\oplus$. The planets are colored by categories in Fig.~\ref{fig:arch}. We see the frequency of super-Earths increasing and sub-Neptunes decreasing with increasing gap complexity in Fig.~\ref{fig:changing}. In Fig.~\ref{fig:sigplot}, super-Earths are more likely to have a high-$C$ while sub-Neptunes are more likely to have a low-$C$.

We calculate the percentage here of each type of planet considering only planets between 1-3 R$_\oplus$ in our systems divided into $C$$<$0.165 and $C$$>$0.35. In low-$C$ systems, 47\% of planets (163 planets) have their median planet radius in the super-Earth regime while 53\% (182) are in the sub-Neptune regime. In high-$C$, 63\% (103) are super-Earths while only 37\% (60) are sub-Neptune. Super-Earths are 1.3 times more prevalent while sub-Neptunes are similarly 1.4 times less prevalent in high-$C$ systems compared to low-$C$ systems.

\section{Discussion} \label{sec:discussion}

We find that the radius distribution of \textit{Kepler} planets in 3+ planet systems is correlated with the spacing of planets within a system. While the high-$C$ systems and low-$C$ systems sample the same stellar properties, they sample different spreads in orbital period. Thus, we discuss if the changes in the radius distribution are a result of period sampling or rather if they could be astrophysical. 

\subsection{Sub-Earth Discussion} \label{sec:subearthdis}

While the gap complexities of systems with sub-Earths for the most part follow the frequency of all of the systems, shown in Fig.~\ref{fig:periods}, there are eight systems with gap complexities between 0.07 and 0.105. Although this is a small number, it is approximately twice the expected amount and results in the sharp increase in the significance of the K-S tests in Fig.~\ref{fig:kstests}.

Furthermore, three of the eight systems have exactly three sub-Earths and three have four or more sub-Earths. Included in these three high-multiplicity sub-Earth systems are KOI-2859 (Kepler-1371) and KOI-4032 (Kepler-1542)---two systems of all 5 planets having median planet radius below that of Earth (see Fig.~\ref{fig:arch}). The 6 systems with 3+ sub-Earths that are all in this narrow region of gap complexity (0.07$<$C$<$0.105) have 23 sub-Earths. Removing these 23 planets accounts for about half of the over-abundance of sub-Earths in low-$C$ systems.

While KOI-2859 and KOI-4032 have non-negligible gap complexities, their average $\Delta P/P_i$ are 0.29 and 0.26 respectively. In Fig.~\ref{fig:allpl}, we see the sub-Earths are primarily under 20 days. To find multiple sub-Earths under 20 days the spacing between planets must be small keeping the planets within 20 days. A tight spacing is correlated with a low gap complexity as shown in Fig.~\ref{fig:periods}.

%Indeed 83\% (37) of sub-Earth hosting systems have average $\Delta P/P_i < 2$ making up a large portion of the systems at the bottom of Fig.~\ref{fig:periods}. CALCULATE FOR LOW GAP COMPLEXITY SUB-EARTH SYSTEMS COMPARED

This brief analysis of specific systems indicates that the increase in sub-Earths at low-$C$ may be an effect of observational bias toward shorter periods for sub-Earth observations. The \textit{Kepler} telescope can only find sub-Earths with a high number of transits thus favoring shorter orbital periods. The Solar System is an example of a system of sub-Earths with a higher $C$ than the sub-Earth systems found by \textit{Kepler}. With an incomplete survey of small planets, we may be missing systems of sub-Earths with high-$C$. 

In addition to missing irregularly spaced systems of sub-Earths, we could also be missing sub-Earths in our already observed high-$C$ systems. \citet{Gilbert2020} presents several lines of evidence that systems of $C$$>$0.33 have unobserved planets. The value of $C$ for an intrinsically 4-planet system that is evenly spaced in period ratio where one interior planet is unobserved is 0.336. These missing interior planets may not be transiting because of higher mutual inclinations (see next paragraph), but \citet{Zhu2018} find that higher intrinsic multiplicity systems tend to have lower mutual inclinations. Our missing sub-Earths in high-$C$ systems could support that these missing interior planets are small and below the detection threshold. However, the lack of sub-Earths begins at a much lower value of $C$ than suggested in \citet{Gilbert2020} for missing interior planets with the deficit most prominent at $C$$>$0.11. 

If \textit{Kepler} high-$C$ systems do host unobserved sub-Earths but they are not transiting, this could point to outer giants disrupting smaller planets in the inner system and raising their mutual inclinations. Raising the inclination and ejecting planetesimals in the region of Mars through the migration of Jupiter and Saturn or an instability amongst the giant planets is a common solution to the ``small Mars problem'' in the Solar System \citep{Morbidelli2016, Clement2018, Clement2019}. As in the early Solar System \citep{Morbidelli2007}, cold giant planets in exoplanet systems should similarly form in resonant chains driven by gravitation torques of the disk. The giant planets then scatter small objects causing the resonant chains to break and further scattering materials as the planets migrate \citep{Malhotra1993,Gomes2005}. High-$C$ systems may represent those with multiple giant planets which have migrated and ejected smaller sub-Earth bodies from the system or raised their mutual inclinations to the point where they are unobserved.

Our discussion in this section would be aided by a future work detailing how $C$ changes with changing the inner planet's period, changing the outer planet's period, and discovering a new planet interior or exterior to the current planets.

\subsection{Super-Earth and Sub-Neptune Discussion} \label{sec:radgapdis}

Unlike the sub-Earths, we find that the change in the abundance of super-Earths and Sub-Neptunes is not easily explained by the periods sampled in low- and high-$C$ systems. In Fig.~\ref{fig:allpl}, we see there is a small number of hot, near Earth-radii planets which for the most part have high gap complexities. However, if we remove all systems which contain either a shorter than 2 day orbit or longer than 365 day orbit our results remain the same. If this restriction is made there are only 41 systems with $C$$>$0.35, making the uncertainties in the frequency much larger, but the over-abundance of super-Earths and under-abundance of sub-Neptunes remains in high-$C$ systems with approximately the same K-S significance when compared to low-$C$.

The division between low- and high-$C$ does correspond with the value of a system of 3-observed planets and one interior unobserved planet (discussed in the Sect.~\ref{sec:subearthdis}. However, it is less clear why the unobserved interior planet would be preferentially a sub-Neptune over a super-Earth. Nearly two-thirds of high-$C$ systems would need a missing sub-Neptune to equalize the amount of super-Earths and sub-Neptunes. High-$C$ systems do represent a larger spread in periods and sub-Neptunes tend to have longer periods than super-Earths, so there may be a slight bias to missing sub-Neptunes but not enough to match the proportions in low-$C$ systems. An injection-recovery test with varied initial distributions or planet radius and intrinsic gap complexities is reserved for a future work.

Our results agree with those of \citet{Goyal2024}. They split \textit{Kepler} multis into ones of only planets less than 1.6 R$_\oplus$ and systems with only planets between 1.6 and 4.0 R$_\oplus$. They find the systems with only smaller planets have a higher uniformity in period ratios measured by a small ($\leq$0.1) Gini index. The Gini index of period ratios has a much larger spread and higher maximum value in the systems of larger planets. \citet{Goyal2024} also finds a greater mass uniformity amongst the larger planets and discusses that the lower spacing uniformity does not fit in with the low-entropy dynamical pathways which are preferred by their other results.

Thus, we too turn our discussion to astrophysical formation mechanisms. If the high-$C$ systems initially had equal amounts of super-Earths and sub-Neptunes, it is possible that the mechanism that raised their gap complexities also changed approximately 10-20\% of sub-Neptunes to super-Earths. High-$C$ systems may represent high entropy formation environments which is further supported by \citet{He2023} finding cold giant planets in high-$C$ systems. In contrast, \citet{Kong2024} find that the dynamical stirring from giant planets preferentially collide the most different planetesimals in mass and spacing leaving the final system more compact and uniform than in a low-entropy formation environment (also supported by \citet{Lammers2023}).

If planets in high-$C$ systems experience more collisions, this lack of sub-Neptunes could be evidence of impact-driven atmosphere loss \citep{Schlichting2015,Lozovsky2023}. The formation simulations of \citet{Dawson2016} show that systems with low solid surface density and less gas during inner system formation tend to produce rockier planets with wider spacings. \citet{Chance2022} analyzes the simulations of \citet{Dawson2016} for the effect of impact-driven atmosphere erosion and finds it is able to reproduce the radius valley although a less empty radius valley than predicted by photoevaporation. Similarly, the formation models of \citet{Izidoro2022} invoke a period of instability amongst the inner planets in resonant configurations which lead to giant impacts. They argue that the giant impacts are efficient at stripping all primordial atmospheres and the radius valley is between water-rich larger planets and rocky smaller planets.

Alternatively, the difference could form earlier with systems that evolve to high-$C$ initially having less volatile-rich planets. The mechanism proposed for changing the radius distribution of S-type planets in binary star systems in \citet{Sullivan2023} and \citet{Sullivan2024} is the binary decreasing the timescale and mass of the disk which then inhibits core formation. Similarly giant planets, proposed to exist in high-$C$ systems, may decrease the inward pebble drift \citep{Lambrechts2014} and decrease the mass available in the protoplanetary disk \citet{Thommes2003, Haghighipour2012}. The decrease of inward pebble drift especially that of volatile-rich material may lead to the overabundance of super-Earths. The shortening of the lifespan of the protoplanetary disk may decrease core masses and leave the planets in more excited orbits which may be unobserved and/or lead to atmosphere erosion by collisions.

The over abundance of super-Earths and under abundance of sub-Neptunes in high-$C$ systems is an effect of relatively small numbers. An additional 10 systems of three sub-Neptunes would change the result. Thus future work should continue to grow the sample of multis and searches should look for missing planets in high-$C$ systems.

\section{Conclusion} \label{sec:conclusion}

Two demographic results that emerged from the numerous exoplanets found by the \textit{Kepler} space telescope are the existence of a radius valley---a dip in the frequency of planet radius separating ``super-Earths'' from ``sub-Neptunes''---and the existence of compact multis---adjacent planets in multiplanet systems have period ratios smaller than those in the Solar System. Both trends provide insights into and constraints on the formation and evolution of exoplanet systems. In this work, we analyze how the frequency of planet radii less than 6 R$_\oplus$ changes with a system's gap complexity---a measure of how equally spaced planets are in a system of three or more planets.

Using a sample of 783 transiting planet candidates in 234 systems of three or more planets around FGK stars, we find that the radius of planets in a system correlates with the system's gap complexity. Specifically, we find:

\begin{itemize}
    \item As gap complexity increases, the frequency of sub-Earths ($<$1 R$_\oplus$) and sub-Neptunes (1.9$<$R$<$3.0 R$_\oplus$) decreases and the frequency of super-Earths (1.0$<$R$<$1.9 R$_\oplus$) increases.
    \item The average gap complexity of planets in the peaks and troughs of the radius distribution differs by approximately 1.5$\sigma$ from the expected value.
    \item Systems with $C$$<$0.105 have 2.5 times more observed sub-Earths than systems above $C$$>$0.11.
    \item  While systems with $C$$<$0.165 have approximately equal numbers of observed super-Earths and sub-Neptunes, systems with $C$$>$0.35 have 1.3 times more super-Earths and 1.4 times fewer sub-Neptunes.
    \item A two-sample K-S test indicates that planets in systems with $C$$<$0.105 and in systems with $C$$>$0.11 are not drawn from the same radius distribution below 6 R$_\oplus$, with a p-value of 0.004. Similarly, planets in systems with $C$$<$0.165 and in systems with $C$$>$0.35 are not drawn from the same radius distribution between 1-3 R$_\oplus$, with a p-value of 0.0008.
\end{itemize}

%The changes in the radius distribution becomes more evident when systems are split into low and high gap complexity samples. A two-sample Kolmogorov-Smirnov (K-S) test confirms that the radii of planets in low-$C$ and high-$C$ systems are drawn from different distributions. The K-S test returns a p-value of 0.0044 for planet radius in systems with $C$$<$0.105 compared to $C$$>$0.11. We find that the difference in the distribution when divided at this value of $C$ is primarily in the sub-Earth regime. Low-$C$ systems below a complexity of 0.105 have 2.5 times more sub-Earths than high-$C$ systems.

%When narrowing our analysis to the region of the radius valley between one and three times the radius of Earth, the K-S test returns an even lower p-value (0.0008) when splitting systems into $C$$<$0.165 and $C$$>$0.35. In this case, the planets in low-$C$ systems exhibit a defined radius valley with approximately equal frequencies of super-Earths and sub-Neptunes. High-$C$ systems above a gap complexity of 0.35 have a suppression of sub-Neptune frequency to below that of planets in the radius valley and an increase in the frequency of super-Earths. In high-$C$ systems, 63\% of planets have a radius below the radius valley (and above the radius of Earth) and 37\% of planets are above the radius valley.

These differences in the radius distribution relative to gap complexity may hold clues to how planetary systems are formed. The higher prevalence of sub-Earths in low-$C$ systems may only be caused by a few systems with multiple sub-Earth and the tight nature of low-$C$ systems. However, if sub-Earth's are not found in high-$C$ systems this could point to outer giant planets raising the inclinations and ejecting sub-Earths from systems. 

The increased frequency of super-Earth and decreased frequency of sub-Neptunes around high-$C$ systems remains even when we make stringent cuts to the maximum and minimum periods. The sub-Neptunes could be experiencing impact-driven atmosphere loss in high-$C$ systems converting them to super-Earths. Alternatively, the planets in high-$C$ systems may be forming from less volatile building blocks than in low-$C$ systems.

Testing if these formation scenarios can reproduce the radius correlation with gap complexity will be important future work. Significant to understanding these formation scenarios will be future searches for giant planets exterior to inner systems such as in \citet{Weiss2024}. A large sample of giant planets exterior to high multiplicity inner systems would allow for the study of how the cold giant planet's mass, eccentricity, and inclination relate to the properties of the inner system such as gap complexity. Lastly, understanding the formation mechanisms that explain our findings is crucial for fitting into the picture the Solar System with its medium gap complexity, rocky planets, cold giant planets, and epoch of giant impacts.

%% Also note that the akcnowlodgment environment does not support long amounts of text. If you have a lot of people and institutions to acknowledge, do not use this command. Instead, create a new \section{Acknowledgments}.
\begin{acknowledgments}
The authors thank the anonymous referee for their feedback which enhanced the quality of this work. We appreciate the pointers and feedback from K. Sullivan. This work was inspired by a number of talks and posters at the Extreme Solar Systems V conference in 2024 including those by L.~Weiss, K.~Sullivan, and B.~Liberles. We additionally appreciate M.~He, A.~Gupta, M.~MacDonald, and the Steffen Research Group for stimulating discussions. We acknowledge support from the Astrophysics Research Center at the Open University of Israel, the College of Sciences at the University of Nevada, Las Vegas, and the Nevada Center for Astrophysics. A.V. acknowledges support from the Israel Science Foundation (ISF) grants 770/21 and 773/21.\\
\end{acknowledgments}


%%\vspace{5mm}
%%\facilities{HST(STIS), Swift(XRT and UVOT), AAVSO, CTIO:1.3m, CTIO:1.5m,CXO}


\software{CMasher \citep{cmasher},  NumPy \citep{numpy}, SciPy \citep{scipy}, Matplotlib \citep{matplotlib}, ChatGPT-4 (Assisted in the writing of the abstract, \citealt{openai})
          }




\bibliography{radgap}{}
\bibliographystyle{aasjournal}



\end{document}